\documentclass[%
aip,
apl,
reprint,
preprintnumbers,
superscriptaddress,
amsfonts,
amssymb,
amsmath
]{revtex4-2}
\usepackage{bm,latexsym,mathrsfs,enumerate}
\usepackage{upgreek}
\usepackage[table,x11names]{xcolor}
\usepackage[breaklinks=true,
unicode=true,
urlcolor = RoyalBlue4,
colorlinks = true,
citecolor = RoyalBlue4,
linkcolor = RoyalBlue4
]{hyperref}
\usepackage{graphicx}
\usepackage{mathtools}
\usepackage[cal=boondoxo,frak=boondox,scr=rsfs]{mathalfa}
\usepackage{savesym}
\usepackage[most]{tcolorbox}
\renewcommand{\vec}[1]{\bm{#1}}

\usepackage[final]{pdfpages}
\usepackage{tikz}
\makeatletter
\AtBeginDocument{\let\LS@rot\@undefined}
\makeatother

\begin{document}
\title{Curvature-driven homogeneous Dzyaloshinskii--Moriya interaction  and emergent weak ferromagnetism in anisotropic antiferromagnetic spin chains}

\date{February 26, 2021}

\author{Oleksandr~V.~Pylypovskyi}
\affiliation{Helmholtz-Zentrum Dresden-Rossendorf e.V., Institute of Ion Beam Physics and Materials Research, 01328 Dresden, Germany}
\affiliation{Kyiv Academic University, 03142 Kyiv, Ukraine}

\author{Yelyzaveta~A.~Borysenko}
\affiliation{Taras Shevchenko National University of Kyiv, 01601 Kyiv, Ukraine}

\author{J\"{u}rgen Fassbender}
\affiliation{Helmholtz-Zentrum Dresden-Rossendorf e.V., Institute of Ion Beam Physics and Materials Research, 01328 Dresden, Germany}
	
\author{Denis~D.~Sheka}
\affiliation{Taras Shevchenko National University of Kyiv, 01601 Kyiv, Ukraine}

\author{Denys~Makarov}
\email{d.makarov@hzdr.de}
\affiliation{Helmholtz-Zentrum Dresden-Rossendorf e.V., Institute of Ion Beam Physics and Materials Research, 01328 Dresden, Germany}

\begin{abstract}
Chiral antiferromagnets are currently considered for broad range of applications in spintronics, spin-orbitronics and magnonics. In contrast to the established approach relying on materials screening, the anisotropic and chiral responses of low-dimensional antifferromagnets can be tailored relying on the geometrical curvature. Here, we consider an achiral, anisotropic antiferromagnetic spin chain and demonstrate that these systems possess geometry-driven effects stemming not only from the exchange interaction but also from the anisotropy. Peculiarly, the anisotropy-driven effects are complementary to the curvature effects stemming from the exchange interaction and rather strong as they are linear in curvature. These effects are responsible for the tilt of the equilibrium direction of vector order parameters and the appearance of the homogeneous Dzyaloshinskii--Moriya interaction. The latter is a source of the geometry-driven weak ferromagnetism emerging in curvilinear antiferromagnetic spin chains. Our findings provide a deeper fundamental insight into the physics of curvilinear antiferromagnets beyond the $\sigma$-model and offer an additional degree of freedom in the design of spintronic and magnonic devices.
\end{abstract}

\maketitle


Antiferromagnets (AFMs) represent a rich class of technologically promising materials, whose magnetic properties are determined by the antiparallel configuration of neighboring spins\cite{Gomonay17,Gomonay18,Baltz18,Liu19}. One of the distinct properties of AFMs is related to the variety of intrinsic crystal symmetries and mechanisms of the exchange\cite{Turov01en} and anisotropy\cite{Turov01en,OGrady20}. Within the phenomenological formalism, this is reflected in the specific energy invariants, mixing components of different vector order parameters\cite{Izyumov84}. In this way, there appear homogeneous Dzyaloshinskii--Moriya (without spatial derivatives of the order parameter) and inhomogeneous Dzyaloshinskii--Moriya  (linear with respect to spatial derivatives of the order parameter) energy terms, which determine the appearance of non-collinear and incommensurable magnetic textures\cite{Izyumov84,Bogdanov89r,Bogdanov02a}. Among them, antiferromagnetic chiral domain walls and skyrmions are perspective information carriers\cite{Gomonay18} and functional elements of prospective devices\cite{Shen19,Shen20}. Microscopically, the local break of the inversion symmetry leads to the Dzyaloshinskii--Moriya interaction (DMI)\cite{Dzyaloshinsky58,Moriya60} and staggered spin-orbit torques\cite{Zelezny14}. This enables an efficient interaction between the electrical current and magnetic textures, resulting in ultrahigh velocities of magnetic solitons\cite{Barker16,Gomonay16a}. In addition to the intrinsic properties of the crystal lattice, the magnetic responses can be tuned by the geometry of the samples, which allows to utilize boundary conditions\cite{Hedrich21,Pylypovskyi21a} and geometrical curvatures to design non-collinear magnetic states\cite{Castillo-Sepulveda17} and dispersion curves\cite{Pylypovskyi20}.

In this work, we develop the analytical approach beyond the $\sigma$-model to describe curvilinear one-dimensional (1D) AFMs and determine conditions when they possess the geometry-driven weak ferromagnetism. We show that in contrast to ferromagnets (FMs), AFMs exhibit the geometry-driven modification of the magnetic responses stemming not only from the exchange, but also from the anisotropy interaction. The key consequences of the fact that the anisotropy axis follows the shape of the spin chain are the appearence of the homogeneous DMI energy term and the tilt of the anisotropy axis in the osculating plane. These effects originate from the multiplicity of magnetic sublattices in AFMs. Therefore, our results are of importance for curvilinear ferrimagnetic systems as well.


We consider intrinsically achiral, anisotropic antiferromagnetic spin chains with the Hamiltonian $\mathscr{H} = \mathscr{H}_\text{x} + \mathscr{H}_\text{a} + \mathscr{H}_\text{f}$, where $\mathscr{H}_\text{x}$ represents the nearest-neighbor exchange, $\mathscr{H}_\text{a}$ is the anisotropic part of the Hamiltonian and $\mathscr{H}_\text{f}$ is the interaction with the external magnetic field $\vec{H}$. The magnetic moments $\vec{M}_i$ with $i = \overline{0,N-1}$ and $N$ being the total number of spins are arranged along a space curve $\vec{\gamma}(s)$ with $s$ being the arc length. The geometrical properties of $\vec{\gamma}$ are determined\cite{Kuehnel15} by the curvature, $\kappa(s) = |\vec{\gamma}'\times \vec{\gamma}''|$, and torsion $\tau(s) = [\vec{\gamma}'\times \vec{\gamma}'']\cdot \vec{\gamma}'''/|\vec{\gamma}''|^2$, where prime indicates the derivative with respect to $s$. The local reference frame is determined by the tangential $\vec{e}_\textsc{t} = \vec{\gamma}'$, normal $\vec{e}_\textsc{n} = \vec{\gamma}''/\kappa$ and binormal $\vec{e}_\textsc{b} = \vec{e}_\textsc{t} \times \vec{e}_\textsc{n}$ directions, respectively. In the following, we discuss weakly curved geometries, i.e., $\kappa, |\tau| \ll 1/\ell \ll 1/a_0$, where $\ell$ is the magnetic length, determined by the competition between the exchange and anisotropy interactions, $a_0$ is the lattice constant. We assume that the system is far below the N\'{e}el temperature and all magnetic moments are of a constant length $M_0 = 2\mu_\textsc{b}S$ with $\mu_\textsc{b}$ being the Bohr magneton and $S$ being the spin length.

\begin{figure}[t]
\includegraphics[width=\linewidth]{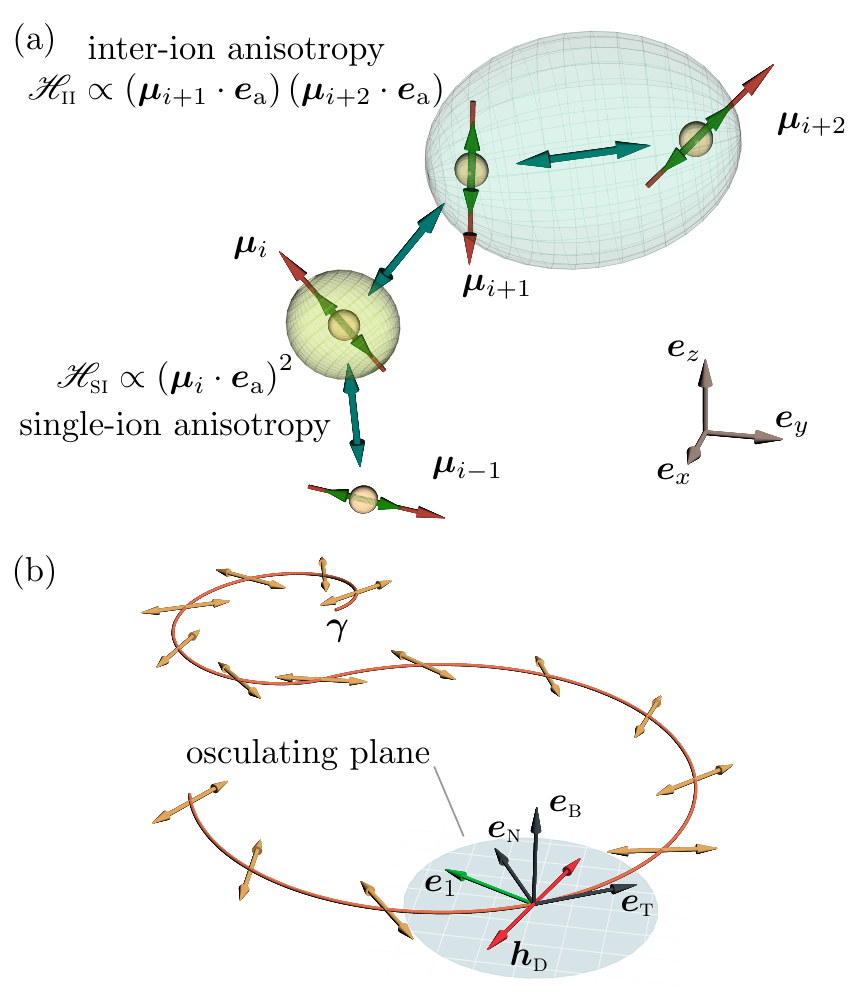}
\caption{\textbf{Anisotropic antiferromagnetic spin chain} (a) Spin chain with unit vectors of magnetic moments $\vec{\mu}_i$ (red arrows). Double arrows (green) show the anisotropy axes of the single- and inter-ion anisotropies for the given magnetic sites $\mathscr{H}_\textsc{si}$ and $\mathscr{H}_\textsc{ii}$, respectively. (b) Micromagnetic representation of the spin chain along the curve $\vec{\gamma}$ with $\vec{n}$ being the N\'{e}el vector. Local TNB reference frame is indicated with black arrows. Vector $\vec{h}_\textsc{d}$ represents the effective field of the longitudinal homogeneous DMI. The arrow $\vec{e}_1$ shows the equilibrium direction of the anisotropy axes due to the interplay between $w_\text{a}$ and $\widetilde{w}_\text{a}$.}
\label{fig:curve}
\end{figure}

The nearest-neighbor antiferromagnetic exchange $\mathscr{H}_\text{x} = (\mathscr{J}S^2/2) \sum_i \vec{\mu}_i\cdot \vec{\mu}_{i+1}$ with the exchange integral $\mathscr{J} > 0$ and $\vec{\mu}_i = \vec{M}_i/M_0$ allows to identify two sublattices of magnetic moments and introduce staggered (N\'{e}el) and ferromagnetic vector order parameters, $\vec{n}_i = (\vec{\mu}_{2i} - \vec{\mu}_{2i+1})/2$ and $\vec{m}_i = (\vec{\mu}_{2i} + \vec{\mu}_{2i+1})/2$, respectively. The micromagnetic exchange energy reads
\begin{equation}\label{eq:exchange-n-m}
\begin{aligned}
E_\text{x} & = \int \! w_\text{x}\, \mathrm{d}s\\
w_\text{x} & = \Lambda m^2 + A_0 (\vec{n}'^2 - \vec{m}'^2) + \lambda \vec{m}\cdot \vec{n}',
\end{aligned}
\end{equation}
where $\vec{n}$ and $\vec{m}$ are the continuum counterparts of $\vec{n}_i$ and $\vec{m}_i$, respectively, the constant of the uniform exchange $\Lambda = 2\mathscr{J}S^2/a_0$, exchange stiffness $A_0 = \mathscr{J}S^2a_0$, the parity breaking coefficient $\lambda = 2\mathscr{J}S^2$, see Supplementary material for details. We note that the exchange stiffness within the $\sigma$-model approach $A = A_0/2$ because of the renormalization due to the parity breaking term\cite{Tveten16}. The expression~\eqref{eq:exchange-n-m} is similar to the exchange energy of 1D AFMs\cite{Tveten16} with the difference that the spatial derivatives are taken in the curvilinear reference frame. The curvature effects stemming from the exchange term $\vec{n}'^2$ are discussed in Ref.\cite{Pylypovskyi20}. The term $\vec{m}'^2$ is expected to affect the system near the spin-flip transition. As we will show in the following, the parity breaking term in~\eqref{eq:exchange-n-m} scales linearly with $\kappa$ and~$\tau$ and is a source of the weak ferromagnetism in curvilinear AFM spin chains. 

The anisotropic contribution to the microscopic Hamiltonian can be presented by the so-called inter-ion and single-ion anisotropies\cite{Ivanov05a}, see Fig.~\ref{fig:curve}(a). In the simplest case of one symmetry axis $\vec{e}_\text{a}$ at the position of $i$-th spin, the anisotropic part of the Hamiltonian reads
\begin{equation}\label{eq:hamiltonian-ani}
\begin{aligned}
\mathscr{H}_\text{a} = & - \dfrac{\mathscr{K}_\textsc{si}S^2}{2} \sum_i (\vec{\mu}_i\cdot \vec{e}_\text{a})^2 \\
& + \dfrac{\mathscr{K}_\textsc{ii}S^2}{2} \sum_i (\vec{\mu}_i\cdot \vec{e}_\text{a})(\vec{\mu}_{i+1}\cdot \vec{e}_\text{a}).
\end{aligned}
\end{equation}
The first term represents the single-ion anisotropy with the constant $\mathscr{K}_\textsc{si}$, which is determined by the spin-orbit interaction and is relevant for $S\ge 1$\cite{Turov01en,Ivanov05a}. The second term determines the inter-ion anisotropic interactions with the anisotropy constant $\mathscr{K}_\textsc{ii}$, originating from the anisotropic exchange interaction, spin-orbit interaction and dipolar interaction\cite{Turov01en,Ivanov05a}. Micromagnetically, single-ion and inter-ion anisotropies have different contributions to the characteristic magnetic fields of the phase transitions\cite{Ivanov05a}. If the dipolar interaction has no other competing anisotropic terms (the case of an isotropic AFM with dipolar interaction), it leads to the hard-axis anisotropy with $\vec{e}_\text{a} = \vec{e}_\textsc{t}$\cite{Pylypovskyi20}.

In the following, we limit ourselves by the case of the tangential direction of the anisotropy axis, $\vec{e}_\text{a} = \vec{e}_\textsc{t}$. In this case, the micromagnetic expression for the anisotropy energy reads
\begin{equation}\label{eq:anisotropy}
\begin{aligned}
E_\text{a} = & \int\!\left( w_\text{a} + w_\text{pb} + \widetilde{w}_\text{a} + w_\textsc{d} \right) \mathrm{d}s, &&  \\
w_\text{a} = & - K_n n_\textsc{t}^2 - K_m m_\textsc{t}^2, & w_\text{pb} & = \lambda_\textsc{t} m_\textsc{t}n_\textsc{t}',\! \\
\widetilde{w}_\text{a} = & - \widetilde{K}_n n_\textsc{t}n_\textsc{n} - \widetilde{K}_m m_\textsc{t}m_\textsc{n}, & w_\textsc{d} & = \vec{h}_\textsc{d} \cdot \vec{m},
\end{aligned}
\end{equation}
Here, the first term $w_\text{a}$ with $K_{n,m} = (\mathscr{K}_\textsc{si} \pm \mathscr{K}_\textsc{ii}) S^2/(2a_0)$ represents the regular micromagnetic anisotropy, also present in straight spin chains with the given easy axis~$\vec{e}_\text{a}$. The term $w_\text{pb}$ in~\eqref{eq:anisotropy} with $\lambda_\textsc{t} = \mathscr{K}_\textsc{ii}S^2$ introduces anisotropy into the exchange-driven parity breaking term. Other terms are determined by the geometrical parameters of $\vec{\gamma}$, see Fig.~\ref{fig:curve}(b).

\begin{figure}
\includegraphics[width=\linewidth]{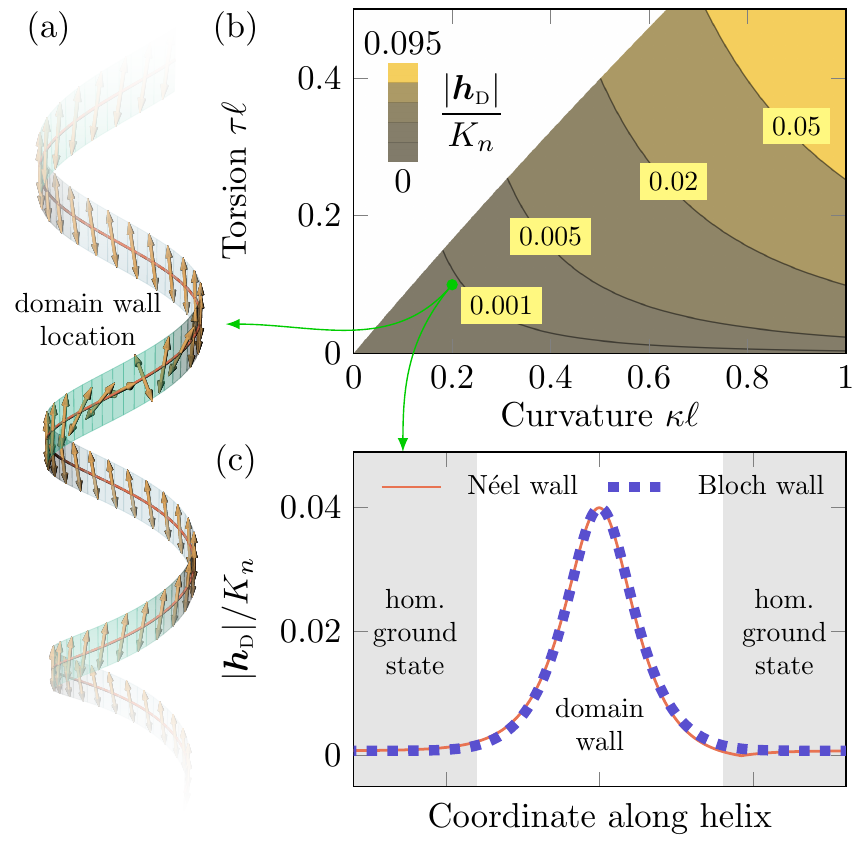}
\caption{\textbf{Curvature-driven homogeneous DMI.} Domain wall in an AFM helix. (b) The homogeneous Dzyaloshinskii field $|\vec{h}_\textsc{d}|$ for the case of the homogeneous ground state\cite{Pylypovskyi20} of the helical spin chain in the absence of the inter-ion anisotropy ($\varsigma = 0.1$). White region corresponds to the periodic ground state\cite{Pylypovskyi20}. (c) $|\vec{h}_\textsc{d}|$ for Bloch and N\'{e}el domain walls on a helical spin chain ($\kappa\ell = 0.2$, $\tau\ell=0.1$, $\varsigma = 0.1$). The region with the ground state of $\vec{n}$ is highlighted in gray.}
\label{fig:curv-dm}
\end{figure}

The emergent homogeneous (longitudinal) DMI term $w_\textsc{d}$ indicates the presence of a weak ferromagnetism-like contribution with the Dzyaloshinskii effective field $\vec{h}_\textsc{d} = D_\textsc{si}(n_\textsc{n}\vec{e}_\textsc{t} + n_\textsc{t}\vec{e}_\textsc{n})$ and $D_\textsc{si} = \kappa \mathscr{K}_\textsc{si}S^2/2$. These energy invariants can be present in crystals with $2_z$ or $\sigma_z$ symmetry if $\vec{e}_z$ axis is associated with the tangential direction\cite{Turov65} and are responsible for the nonlinearity of AFM resonance\cite{Gufan74}. The contribution of this term can be expected for any curve with $\kappa \neq 0$ if the AFM texture possesses the tangential and (or) normal components of $\vec{n}$. In flat systems $\vec{n} = \vec{e}_\textsc{b}$, which leads to $\vec{h}_\textsc{d} = 0$. In contrast, in space curves the normal and binormal components of $\vec{n}$ are present, e.g., in the homogeneous ground state of an AFM helix\cite{Pylypovskyi20}, see Fig.~\ref{fig:curv-dm}(a). The strength of $\vec{h}_\textsc{d}$ increases with torsion and curvature and can reach up to about $10\%$ of the anisotropy field [Fig.~\ref{fig:curv-dm}(b)]. Alternatively, the homogeneous DMI field can be enhanced by orders of magnitude at the location non-collinear AFM textures, such as Bloch or N\'{e}el domain walls, see Fig.~\ref{fig:curv-dm}(c). We note, that this enhancement of the $\vec{h}_\textsc{d}$ at the location of non-collinear textures is due to the non-zero magnetization of the domain wall in a 1D system\cite{Papanicolaou95a,Tveten16}. The finite magnetization at the texture location makes the contribution of the $w_\textsc{d}$ term significant.

\begin{figure}
\includegraphics[width=\linewidth]{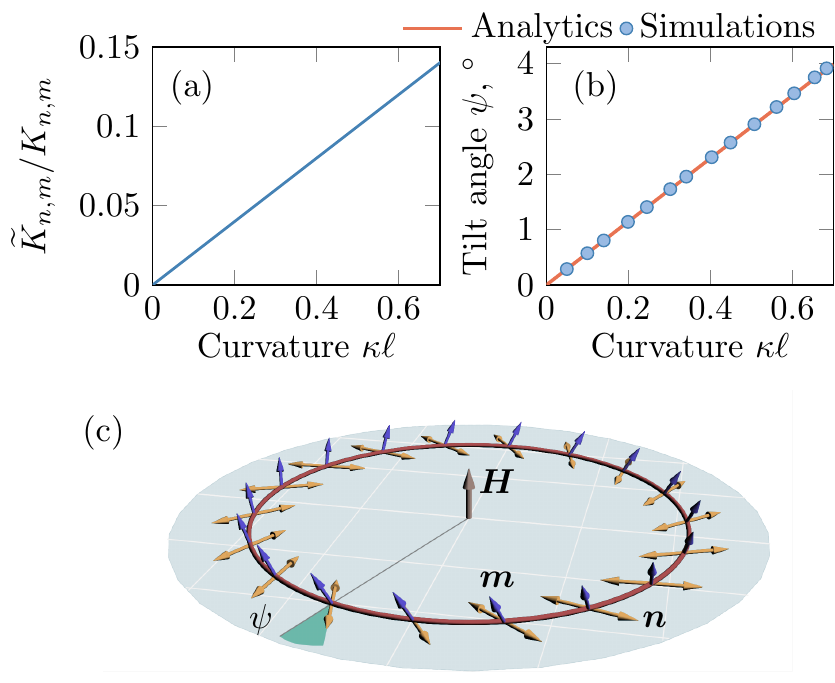}
\caption{\textbf{Curvature-driven tilt of the anisotropy axis.} (a) Relative strength of the geometry-driven anisotropic term, which is proportional to the curvature ($\mathscr{K}_\textsc{ii}/\mathscr{K}_\textsc{si} = 0.3$, $\varsigma = 0.1$). (b) Tilt angle of the anisotropy axis as a function of curvature ($\mathscr{K}_\textsc{ii} = 0$, $\varsigma = 0.1$). (c) AFM ring exposed to an external magnetic field $\vec{H}$ applied perpendicular to the ring plane. In equilibrium, the direction $\psi$ of $\vec{n}$ coincides with the tilted anisotropy axis shown in panel (b).}
\label{fig:curv-ani}
\end{figure}

The term $\widetilde{w}_\text{a}$ with $\widetilde{K}_{n,m} = \kappa (\mathscr{K}_\textsc{si} \pm \mathscr{K}_\textsc{ii})S^2/2$ represents the non-diagonal components of the total anisotropy tensor contributing to the energy in addition to the geometry-driven anisotropy stemming from the exchange interaction. Note, that the latter one is scaled as $\kappa\tau$ and absent in flat curves\cite{Pylypovskyi20} in contrast to $\widetilde{w}_\text{a} \propto \kappa$, see Fig.~\ref{fig:curv-ani}(a). The equilibrium direction of the anisotropy axes is determined by the diagonal form of the anisotropy tensor. The tilt $\psi$ of the easy axis from $\vec{e}_\textsc{n}$ in the case of an AFM ring is shown in Fig.~\ref{fig:curv-ani}(b).

The interaction of the spin chain with a uniform external magnetic field $\vec{H}$ is described by $\mathscr{H}_\text{f} = -M_0 \sum_i \vec{\mu}_i\cdot \vec{H}$. The macroscopic Lagrangian, allowing to determine the equations of motion for the vector order parameters reads 
$L = - 2M_\textsc{s}/\gamma_0 \int \vec{m} \cdot [\vec{n}\times\dot{\vec{n}}]\mathrm{d}s - E_\text{tot}$ with the total energy $E_\text{tot}  =  E_\text{x} + E_\text{a} + E_\text{f} = \int w_\text{tot} \mathrm{d}s$,
where $\gamma_0$ is the gyromagnetic ratio, the dot indicates the derivative with respect to time, $M_\textsc{s} = M_0/(2a_0)$ is the saturation magnetization of one sublattice and $E_\text{f} = -2M_\textsc{s} \int \vec{m}\cdot \vec{H}\mathrm{d}s$ is the energy of the interaction with the magnetic field.  We limit the following discussion by the case of a strong intrinsic anisotropy with the hard axis along $\vec{e}_\textsc{t}$. Then, the magnetic length reads $\ell = \sqrt{A_0/(2|K_n|)}$, where $K_n < 0$. For the magnetic fields smaller than or comparable to the spin-flop field, the magnetization of the ring is small enough, which allows to apply a standard variational approach for the Lagrangian
to determine the relation between $\vec{n}$ and $\vec{m}$ in equilibrium. Taking into account that $\vec{m}\cdot \vec{n} = 0$ and $n^2 + m^2 = 1$, the magnetization reads 
\begin{equation}\label{eq:magnetization}
\vec{m} \approx - \varsigma \vec{n} \times [\omega_0 \dot{\vec{n}} + (\ell\vec{n}' - \vec{H}/H_0)\times \vec{n}],
\end{equation}
where $\varsigma = a_0/(2\ell)$ is the expansion parameter measuring the strength of the effective anisotropy field with respect to the exchange field, $H_0 = \sqrt{\Lambda K_n}/M_\textsc{s}$ is the characteristic magnetic field and $\omega_0 = \gamma_0 H_0$ is the characteristic frequency. The term
\begin{equation}\label{eq:parbreak}
\vec{n}' = (n_\textsc{t}' - \kappa n_\textsc{n})\vec{e}_\textsc{t} + (n_\textsc{n}' + \kappa n_\textsc{t} - \tau n_\textsc{b})\vec{e}_\textsc{n} + (n_\textsc{b}' + \tau n_\textsc{n})\vec{e}_\textsc{b}
\end{equation}
in~\eqref{eq:magnetization} originates from the parity breaking term in the expression for the exchange energy~\eqref{eq:exchange-n-m}. This is a source of the weak ferromagnetism in curvilinear AFM spin chains, which is manifested by the presence of a non-zero radial magnetization. Based on~\eqref{eq:magnetization}, for the weakly curved spin chains supporting homogeneous textures in the TNB reference frame, the magnetization reads $\vec{m}= \varsigma\ell \left[\kappa n_\textsc{n}\vec{e}_\textsc{t} + (\tau n_\textsc{b} - \kappa n_\textsc{t})\vec{e}_\textsc{n} - \tau n_\textsc{n}\vec{e}_\textsc{b}\right]$. Therefore, for any spin chain arranged along either a planar curve $\vec{\gamma}$ with $\vec{n} \neq \vec{e}_\textsc{b}$ or a space curve, the system will reveal a weak ferromagnetic response. The strength of the weak ferromagnetism is determined 
by the curvature and torsion. 

The discussed here curvature effects from anisotropy are weaker than those from exchange. Therefore, to explore them, in the following we identify a system, where these effects are pronounced and drive the response of the system. For instance, it is insightful to consider a ring geometry $\vec{\gamma} = R \{ \cos(s/R), \sin(s/R), 0 \}$, which implies constant curvature $\kappa = 1/R$ and $\tau = 0$. In this case, the exchange-driven chiral effects are not active and the ground state $\vec{n}= \vec{e}_\textsc{b}$ is determined by the exchange-driven easy-axis anisotropy  with the coefficient $K_\text{x} = \kappa^2A$\cite{Pylypovskyi20}.
Being exposed to a magnetic field applied along the ring axis, $\vec{H} = H\vec{e}_z$, the spin chain experiences the spin-flop transition at the spin-flop field $H_\text{sf} = \kappa \sqrt{\Lambda A}/M_\textsc{s}$. This leads to the reorientation of the $\vec{n}$ to the direction, which is perpendicular to the external field, see Fig.~\ref{fig:curv-ani}(c). The energy of such a planar configuration of the N\'{e}el vector $\vec{n} = \cos\phi\vec{e}_\textsc{t} + \sin\phi\vec{e}_\textsc{n}$
\begin{equation}\label{eq:norm-state-density}
\dfrac{w_\text{tot}}{K_n} = \underbrace{\ell^2(\kappa + \phi')^2}_{w_\text{x}} + \underbrace{\cos^2\phi}_{w_\text{a}} + \underbrace{2\varsigma\kappa\ell \sin\phi\cos\phi}_{\widetilde{w}_\text{a}} - \dfrac{H^2}{H_0^2}.
\end{equation}
The equilibrium  direction of the N\'{e}el vector in the spin-flop phase is determined by the competition between the contributions of the intrinsic hard axis anisotropy oriented along the tangential direction in $w_\text{a}$ and the curvature-driven non-diagonal term $\widetilde{w}_\text{a}$. This is the curvature effect stemming from the anisotropy interaction. The minimization of $E_\text{tot}$ with respect to $\phi(s)$ leads to a uniform, tilted texture in the local reference frame $\phi = \pm \pi/2 + \psi$, where the tilt angle reads
\begin{equation}\label{eq:psi}
\psi \approx \varsigma\kappa\ell,\quad \kappa\ell \ll 1. 
\end{equation}
A comparison of the theoretical prediction~\eqref{eq:psi} with spin-lattice simulations using in-house developed code SLaSi\cite{SLaSi} is shown in Fig.~\ref{fig:curv-ani}(b), see Supplementary material for details. The equilibrium direction of the magnetization $\vec{m}$ according to~\eqref{eq:magnetization} reads 
\begin{equation}\label{eq:m}
\vec{m} \approx \varsigma  \left( \dfrac{\vec{H}}{H_0} \pm \kappa\ell\vec{e}_\textsc{t} \right).
\end{equation} 
The tilt of magnetization in the tangential direction is the consequence of the parity breaking term in~\eqref{eq:exchange-n-m}. Different signs in~\eqref{eq:m} correspond to the energetically degenerate vortex states of the fundamental homotopy group $\pi_1(\mathbb{S}^1)$ with the opposite chiralities (clockwise and counter-clockwise).

\begin{table}[t]
\caption{\textbf{Comparison of geometry-driven responses in FM and AFM spin chains.} The exchange interaction contributes to the geometry-driven anisotropy and inhomogeneous DMI in both, FMs and AFMs. A variety of sources of anisotropy and multiple vector order parameters lead to the appearance of the geometry-driven anisotropic response and homogeneous DMI of longitudinal symmetry in AFM spin chains. Curvature-driven effects stemming from the anisotropy interaction are absent in FM spin chains.}
\label{tab:comparison}
\includegraphics[width=\linewidth]{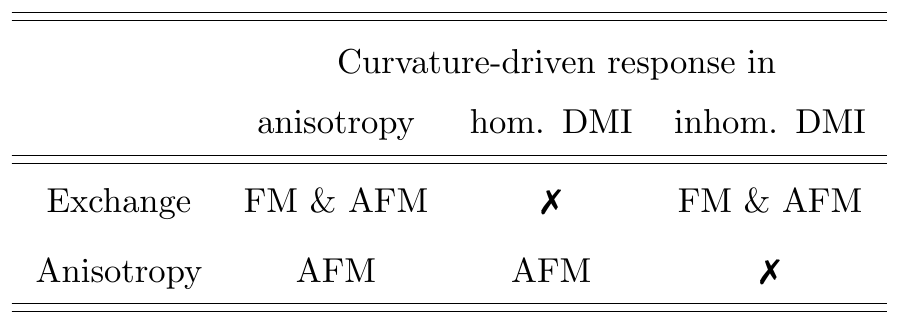}
\end{table}

It is instructive to compare modification of the geometry-driven responses in FMs and AFMs, see Table~\ref{tab:comparison}. The geometry-driven chiral and anisotropic responses for the case of intrinsically achiral FM spin chains originate from the exchange interaction\cite{Sheka15}. They lead to the tilt of the principal axes of anisotropy\cite{Pylypovskyi16,Volkov18} and chiral responses described by the energy invariants in the form of the inhomogeneous DMI \cite{Sheka15c,Pylypovskyi16,Volkov18}. In the intrinsically achiral AFM spin chains, the chiral helimagnetism originates from the nearest-neighbor exchange\cite{Pylypovskyi20}. The interaction, which is tracking the sample geometry in isotropic FMs is the magnetostatics leading to the uniaxial anisotropy with the easy axis along the tangential direction $\vec{e}_{\textsc{t}}$\cite{Slastikov12}. In AFMs, the dipolar interaction leads to the hard axis $\vec{e}_{\textsc{t}}$\cite{Pylypovskyi20}, which makes other anisotropic contributions (even if they are weak) to the Hamiltonian pronounced. The latter leads to the appearence of the anisotropic and weakly ferromagnetic responses in curvilinear AFMs, stemming from the single- and inter-ion anisotropies on the level of the spin Hamiltonian.


In summary, we develop an approach beyond the $\sigma$-model to analyse curvilinear antiferromagnetism in the intrinsically achiral anisotropic spin chains. We identify the conditions of the presence of the weak ferromagnetism in a curvilinear AFM spin chain and determine the contribution to the geometry-driven magnetic responses stemming from the anisotropy interaction, which are specific to AFM materials. The latter emerges due to the presence of two vector order parameters (ferromagnetic and staggered one) in contrast to FMs and is determined by the local curvature $\kappa$ of the chain. Thus, the geometry-driven effects stemming from the anisotropy interaction should be pronounced even for flat curvilinear antiferro- and ferrimagnetic architectures. Considering two general microscopic models of anisotropy, namely, single- and inter-ion anisotropies where the anisotropy hard axis is assumed to be along the tangential direction, we quantify the tilt of the principal axes of anisotropy, identify the appearance of the geometry-driven homogeneous DMI energy term and the additional contribution to the parity breaking coefficient. The physical consequences of the discussed effects are illustrated by the description of the magnetic state of an AFM ring exposed to the external magnetic field. These work opens up an additional route to control AFM textures in spintronic, spin-orbitronics and magnonic devices relying on the geometrical curvature.

See supplementary material for details of analytical derivation of the micromagnetic model from spin-lattice Hamiltonian, spin-lattice simulations and notes on the intrinsic magnetization of curvilinear AFM spin chains.

We thank Dr.~Kostiantyn~V.~Yershov (IFW Dresden) for fruitful discussions. Spin-lattice simulations were made at the Cluster of Taras Shevchenko National University of Kyiv\cite{unicc}. This work was financed in part via the German Research Foundation (DFG) grants MA 5144/22-1, MC 9/22-1, MA 5144/24-1, Alexander von Humboldt Foundation (Research Group Linkage Programme), and by the Ministry of Education and Science of Ukraine (project 19BF052-01) and the National Research Foundation of Ukraine grant (2020.02/0051).

Data availability. The data that support the findings of this study are available from the corresponding author upon reasonable request.
	 
%



\section*{Supplementary material}

# Supplementary material for "Curvature-driven homogeneous Dzyaloshinskii–Moriya interaction and emergent weak ferromagnetism in anisotropic antiferromagnetic spin chains"

Oleksandr V. Pylypovskyi,[1,2] Yelyzaveta A. Borysenko,[3] Jürgen Fassbender,[1] Denis D. Sheka,[3] and Denys Makarov[1, *]

[1]*Helmholtz-Zentrum Dresden-Rossendorf e.V., Institute of Ion Beam Physics and Materials Research, 01328 Dresden, Germany*
[2]*Kyiv Academic University, 03142 Kyiv, Ukraine*
[3]*Taras Shevchenko National University of Kyiv, 01601 Kyiv, Ukraine*

## I. MODEL

We consider a spin chain with magnetic sites lying on a space curve $\gamma(r)$. By introduction of the arc length $s$ parametrization along the $\gamma$, we define the TNB reference frame as

$$e_\mathrm{T} = \gamma', \quad e_\mathrm{N} = \frac{e_\mathrm{T}'}{|e_\mathrm{T}'|}, \quad e_\mathrm{B} = e_\mathrm{T} \times e_\mathrm{N}, \tag{S1}$$

where prime indicates the derivative with respect to $s$. The differential properties of the vectors (S1) are determined by the Frenet–Serret formulas

$$e_\alpha' = \mathscr{F}_{\alpha\beta} e_\beta, \quad \|\mathscr{F}_{\alpha\beta}\| = \begin{pmatrix} 0 & \kappa & 0 \\ -\kappa & 0 & \tau \\ 0 & -\tau & 0 \end{pmatrix}, \tag{S2}$$

where $\kappa(s)$ and $\tau(s)$ are the curvature and torsion of $\gamma(s)$, respectively. If $\gamma(s) \in \mathscr{C}^3(\mathbb{R})$, then $\kappa = |\gamma' \times \gamma''|$ and $\tau = [\gamma' \times \gamma''] \cdot \gamma'''/|\gamma''|^2$ for space curves and $\kappa = \pm|\gamma' \times \gamma''|$ and $\tau = 0$ for planar curves.

In our model, the Hamiltonian of the spin chain contains the exchange and anisotropic contributions. The exchange part reads

$$\mathscr{H}_\mathrm{x} = \mathscr{J} S^2 \sum_i \boldsymbol{\mu}_i \cdot \boldsymbol{\mu}_{i+1} = \mathscr{J} S^2 \sum_i (\boldsymbol{\mu}_{ai} \cdot \boldsymbol{\mu}_{bi} + \boldsymbol{\mu}_{bi} \cdot \boldsymbol{\mu}_{ai+1}), \tag{S3}$$

where $\mathscr{J} > 0$ is the exchange integral, $\boldsymbol{\mu}_i$ is the unit vector of the magnetic moment at $i$-th site, $S$ is the spin length, the distance between neighboring spins is $a_0$ (lattice constant). We assume that the sites at which the spins are located can be divided into two groups, which correspond to two sublattices ($a$ and $b$). This is equivalent to the introduction of dimers (pairs of spins) [1], numerated with $i$. We define vectors of the total and staggered magnetization (Néel vector), normalized by the maximal magnetic moment of the antiferromagnetic (AFM) unit cell consisting of two spins for each of the dimer. The validity of the model (S3) is discussed in [2, 3], see also Chap. 3 in. [4]. In a general case, the anisotropic contribution is given by the single-ion

$$\mathscr{H}_\mathrm{SI} = -\frac{\mathscr{K}_\mathrm{SI} S^2}{2} \sum_i \left[ (\boldsymbol{\mu}_{ai} \cdot \boldsymbol{e}_{ai})^2 + (\boldsymbol{\mu}_{bi} \cdot \boldsymbol{e}_{bi})^2 \right], \tag{S4a}$$

and inter-ion anisotropies

$$\mathscr{H}_\mathrm{II} = \frac{\mathscr{K}_\mathrm{II} S^2}{2} \sum_i \left[ (\boldsymbol{\mu}_{ai} \cdot \boldsymbol{\zeta}_i^i)(\boldsymbol{\mu}_{bi} \cdot \boldsymbol{\zeta}_i^i) + (\boldsymbol{\mu}_{bi} \cdot \boldsymbol{\zeta}_i^{i+1})(\boldsymbol{\mu}_{ai+1} \cdot \boldsymbol{\zeta}_i^{i+1}) \right], \tag{S4b}$$

where $\mathscr{K}_\mathrm{SI}$ and $\mathscr{K}_\mathrm{II}$ are the single-ion and inter-ion anisotropy constants, respectively, $e_{(a,b)i}$ is the anisotropy axis for the $i$-th spin of the sublattice $a$ or $b$, $\boldsymbol{\zeta}_i^i$ is the unit vector in the direction between $i$-th pair of $a$ and $b$, and $\boldsymbol{\zeta}_i^{i+1}$ is the unit vector in the direction between the $i$-th site of $b$ and $(i+1)$-th site of $a$. In the following, we assume that the anisotropy vectors are given by the tangential direction $e_\mathrm{T}$ at a given site. The characteristic length scale in the system is

$$\ell = a_0 \sqrt{\frac{\mathscr{J}}{|\mathscr{K}|}}, \quad \mathscr{K} = \mathscr{K}_\mathrm{SI}, \mathscr{K}_\mathrm{II}. \tag{S5}$$

In the following, we use $\varkappa = \kappa\ell$ and $\sigma = \tau\ell$ for the dimensionless curvature and torsion, respectively. To describe the antiferromagnetically coupled spin chain, we introduce vectors of the total and staggered magnetization (Néel vector), normalized by the maximal magnetic moment of the AFM unit cell consisting of two spins:

$$\boldsymbol{m}_i = \frac{\boldsymbol{\mu}_{ai} + \boldsymbol{\mu}_{bi}}{2}, \quad \boldsymbol{n}_i = \frac{\boldsymbol{\mu}_{ai} - \boldsymbol{\mu}_{bi}}{2}. \tag{S6}$$

---

## II. CONTINUUM TRANSITION

### A. Exchange interaction

Equation (S3) can be rewritten in the following form:

$$\begin{aligned}\mathcal{H}_{\mathrm{x}} &= \mathcal{J} S^2 \sum_i \left(m_i^2 - n_i^2 + \boldsymbol{m}_i \cdot \boldsymbol{m}_{i+1} - \boldsymbol{n}_i \cdot \boldsymbol{n}_{i+1} + \boldsymbol{m}_i \boldsymbol{n}_{i+1} - \boldsymbol{n}_i \boldsymbol{m}_{i+1}\right) \\ &= \frac{\mathcal{J} S^2}{2} \sum_i \left(3 m_i^2 + m_{i+1}^2 - 3 n_i^2 - n_{i+1}^2 - \Delta \boldsymbol{m}_i^2 + \Delta \boldsymbol{n}_i^2 + 2 \boldsymbol{m}_i \cdot \Delta \boldsymbol{n}_i - 2 \boldsymbol{n}_i \cdot \Delta \boldsymbol{m}_i\right),\end{aligned} \quad (S7)$$

where $\Delta \boldsymbol{m}_i = \boldsymbol{m}_{i+1} - \boldsymbol{m}_i$ and $\Delta \boldsymbol{n}_i = \boldsymbol{n}_{i+1} - \boldsymbol{n}_i$. By neglecting the boundary terms, we can assume that $\sum_i m_i^2 = \sum_i m_{i+1}^2$. Taking into account that $m_i^2 + n_i^2 = 1$, we can rewrite (S7) as follows

$$\mathcal{H}_{\mathrm{x}} = \frac{\mathcal{J} S^2}{2} \sum_i \left(8 m_i^2 - \Delta \boldsymbol{m}_i^2 + \Delta \boldsymbol{n}_i^2 + 2 \boldsymbol{m}_i \cdot \Delta \boldsymbol{n}_i - 2 \boldsymbol{n}_i \cdot \Delta \boldsymbol{m}_i\right), \quad (S8)$$

Now, we replace discrete differences by derivatives and sums by integrals as following:

$$\Delta f_i \to 2 a_0 f'(s), \quad \sum_i f_i \to \frac{1}{2 a_0} \int f(s) \mathrm{d}s \quad (S9)$$

with prime indicating the derivative with respect to $s$. Then, the exchange energy reads

$$E_{\mathrm{x}} = \int \mathrm{d}s \left[\Lambda m^2 + A_0 (\boldsymbol{n}'^2 - \boldsymbol{m}'^2) + \lambda \boldsymbol{m} \cdot \boldsymbol{n}'\right], \quad (1)$$

with

$$\Lambda = 2 \frac{\mathcal{J} S^2}{a_0}, \quad A_0 = \mathcal{J} S^2 a_0, \quad \lambda = 2 \mathcal{J} S^2. \quad (S10)$$

and constraints

$$\boldsymbol{m} \cdot \boldsymbol{n} = 0, \quad m^2 + n^2 = 1. \quad (S11)$$

### B. Single-ion anisotropy

In the continuum representation, the Hamiltonian (S4a) reads

$$\begin{aligned}\mathcal{H}_{\mathrm{SI}} = -\frac{\mathcal{K}_{\mathrm{SI}} S^2}{2} \sum_i &\left[(\boldsymbol{m}_i \cdot \boldsymbol{e}_{ai})^2 + (\boldsymbol{n}_i \cdot \boldsymbol{e}_{ai})^2 + 2 (\boldsymbol{m}_i \cdot \boldsymbol{e}_{ai})(\boldsymbol{n}_i \cdot \boldsymbol{e}_{ai})\right. \\ &\left.+ (\boldsymbol{m}_i \cdot \boldsymbol{e}_{bi})^2 + (\boldsymbol{n}_i \cdot \boldsymbol{e}_{bi})^2 - 2 (\boldsymbol{m}_i \cdot \boldsymbol{e}_{bi})(\boldsymbol{n}_i \cdot \boldsymbol{e}_{bi})\right].\end{aligned} \quad (S12)$$

We take into account that $\boldsymbol{e}_{ai} = \boldsymbol{e}_{\mathrm{T}}(s_i) \equiv \boldsymbol{e}_{\mathrm{T}i}$ and $\boldsymbol{e}_{bi}$ can be expressed in the vicinity of $s_i'$ as

$$\boldsymbol{e}_{bi} \approx \boldsymbol{e}_{\mathrm{T}i} \left(1 - \frac{\kappa^2 a_0^2}{2}\right) + \boldsymbol{e}_{\mathrm{N}i} \left(\kappa a_0 + \frac{a_0^2 \kappa'}{2}\right) + \boldsymbol{e}_{\mathrm{B}i} \frac{\kappa \tau a_0^2}{2}. \quad (S13)$$

In the following, we assume that the curvature, torsion and their derivatives are small, i.e. $\kappa, |\tau| \ll 1/\ell \ll 1/a_0$. Therefore, $\boldsymbol{e}_{\mathrm{T}}(s') \approx \boldsymbol{e}_{\mathrm{T}}(s) + \kappa a_0 \boldsymbol{e}_{\mathrm{N}}(s)$. Then, (S4a) reads

$$\begin{aligned}\mathcal{H}_{\mathrm{SI}} = -\mathcal{K}_{\mathrm{SI}} S^2 \sum_i &\left[m_{\mathrm{T}i}^2 + n_{\mathrm{T}i}^2 + \left(\kappa a_0 + \frac{\kappa' a_0^2}{2}\right)(m_{\mathrm{T}i} m_{\mathrm{N}i} + n_{\mathrm{T}i} n_{\mathrm{N}i} - m_{\mathrm{N}i} n_{\mathrm{T}i} - m_{\mathrm{T}i} n_{\mathrm{N}i})\right. \\ &\left.- \frac{\kappa^2 a_0^2}{2}(m_{\mathrm{T}i}^2 - m_{\mathrm{N}i}^2 + n_{\mathrm{T}i}^2 - n_{\mathrm{N}i}^2 + 2 m_{\mathrm{N}i} n_{\mathrm{N}i} - 2 m_{\mathrm{T}i} n_{\mathrm{T}i}) + \frac{\kappa \tau a_0^2}{2}(m_{\mathrm{T}i} m_{\mathrm{N}i} + n_{\mathrm{T}i} n_{\mathrm{N}i} - m_{\mathrm{N}i} n_{\mathrm{T}i} - m_{\mathrm{T}i} n_{\mathrm{N}i})\right].\end{aligned} \quad (S14)$$

| Quantity | $\Lambda$ | $A_0$ | $\lambda$ | $K_{s1}$ | $K_{i1}$ | $K_{i2}$ | $K_{s2}=D_{\text{SI}}$ | $\lambda_{\text{T}}$ |
|---|---|---|---|---|---|---|---|---|
| Value | $2\dfrac{\mathscr{J}S^2}{a_0}$ | $\mathscr{J}S^2 a_0$ | $2\mathscr{J}S^2$ | $\dfrac{\mathscr{K}_{\text{SI}}S^2}{2a_0}$ | $\dfrac{\mathscr{K}_{\text{II}}S^2}{2a_0}$ | $\kappa\dfrac{\mathscr{K}_{\text{II}}S^2}{2}$ | $\kappa\dfrac{\mathscr{K}_{\text{SI}}S^2}{2}$ | $\mathscr{K}_{\text{II}}S^2$ |
| Order of energy density ($\Lambda$) | 1 | $\left(\dfrac{a_0}{\ell}\right)^2$ | $\dfrac{a_0}{\ell}$ | $\left(\dfrac{a_0}{\ell}\right)^2$ | $\left(\dfrac{a_0}{\ell}\right)^2$ | $\varkappa\left(\dfrac{a_0}{\ell}\right)^3$ | $\varkappa\left(\dfrac{a_0}{\ell}\right)^3$ | $\left(\dfrac{a_0}{\ell}\right)^3$ |
| Order of energy density ($A_0$) | $\left(\dfrac{\ell}{a_0}\right)^2$ | 1 | $\dfrac{\ell}{a_0}$ | 1 | 1 | $\varkappa\dfrac{a_0}{\ell}$ | $\varkappa\dfrac{a_0}{\ell}$ | $\dfrac{a_0}{\ell}$ |

TABLE S1: Order of scales for the micromagnetic units.

Using (S9), we obtain

$$E_a^{\text{SI}} = -K_{s1}\int\left[m_{\text{T}}(s)^2 + n_{\text{T}}(s)^2\right]\text{d}s - K_{s2}\int(m_{\text{T}}m_{\text{N}} + n_{\text{T}}n_{\text{N}})\,\text{d}s + D_{\text{SI}}\int(m_{\text{T}}n_{\text{N}} + m_{\text{N}}n_{\text{T}})\,\text{d}s, \tag{S15}$$

where

$$K_{s1} = \frac{\mathscr{K}_{\text{SI}}S^2}{2a_0}, \quad K_{s2} = D_{\text{SI}} = \kappa\frac{\mathscr{K}_{\text{SI}}S^2}{2} \sim \varkappa\frac{a_0}{\ell}. \tag{S16}$$

Other terms, which are $\propto \kappa' a_0^2$, $\kappa\tau a_0^2$ and $\kappa^2 a_0^2$, are of order $\frac{a_0^2}{\ell^2}$ and will be neglected below.

### C. Inter-ion anisotropy

Taking into account only linear terms in $a_0$, the vectors $\zeta_i^{i,i+1}$ coincide with the corresponding tangential directions along lattice sites. Then, (S4b) reads

$$\mathscr{H}_{\text{II}} = \frac{\mathscr{K}_{\text{II}}S^2}{2}\sum_i\left[2m_{\text{T}i}^2 + 2\kappa a_0(m_{\text{T}i}m_{\text{N}i} - n_{\text{T}i}n_{\text{N}i}) - 2n_{\text{T}i}^2 + (m_{\text{T}i} - n_{\text{T}i})(\Delta m_{\text{T}i} + \Delta n_{\text{T}i})\right. \tag{S17}$$
$$\left. + \kappa a_0(m_{\text{T}i} - n_{\text{T}i})(\Delta m_{\text{N}i} + \Delta n_{\text{N}i}) + \kappa a_0(m_{\text{N}i} - n_{\text{N}i})(\Delta m_{\text{T}i} + \Delta n_{\text{T}i})\right].$$

Using (S9) and omitting terms, which are linear in $a_0$, we obtain

$$E_a^{\text{II}} = -K_{i1}\int(n_{\text{T}}^2 - m_{\text{T}}^2)\,\text{d}s - K_{i2}\int(n_{\text{T}}n_{\text{N}} - m_{\text{T}}m_{\text{N}})\,\text{d}s + \lambda_{\text{T}}\int m_{\text{T}}n_{\text{T}}'\,\text{d}s, \tag{S18}$$

where

$$K_{i1} = \frac{\mathscr{K}_{\text{II}}S^2}{2a_0}, \quad K_{i2} = \kappa\frac{\mathscr{K}_{\text{II}}S^2}{2} \sim \varkappa\frac{a_0}{\ell}, \quad \lambda_{\text{T}} = \mathscr{K}_{\text{II}}S^2 \sim \frac{a_0}{\ell}. \tag{S19}$$

### III. LAGRANGIAN OF THE ANTIFERROMAGNETIC SPIN CHAIN AND THE $\sigma$-MODEL

The density of the effective Lagrangian $L = \int\mathscr{L}\,\text{d}s$ can be written as:

$$\mathscr{L} = -\frac{2M_{\text{S}}}{\gamma_0}\boldsymbol{m}\cdot[\boldsymbol{n}\times\dot{\boldsymbol{n}}] - \mathscr{W}, \tag{S20}$$
$$\mathscr{W} = \Lambda\boldsymbol{m}^2 + A_0(\boldsymbol{n}'^2 - \boldsymbol{m}'^2) + \boldsymbol{n}'\hat{\lambda}\boldsymbol{m} - \boldsymbol{n}\hat{K}_n\boldsymbol{n} - \boldsymbol{m}\hat{K}_m\boldsymbol{m} + (\boldsymbol{h}_{\text{D}} - 2M_{\text{S}}\boldsymbol{H})\cdot\boldsymbol{m},$$

where $\gamma_0 = g\mu_{\text{B}}/\hbar$ is the gyromagnetic ratio with $g = 2$ being the Landé $g$-factor, $\mu_{\text{B}}$ being the Bohr magneton and $\hbar$ being the reduced Planck's constant, $\boldsymbol{h}_{\text{D}} = D_{\text{SI}}(n_{\text{N}}\boldsymbol{e}_{\text{T}} + n_{\text{T}}\boldsymbol{e}_{\text{N}})$ is the effective Dzyaloshinskii field, $M_{\text{S}} = g\mu_{\text{B}}S/(2a_0)$ is the linear magnetization of one sublattice, $\boldsymbol{H}$ is the external magnetic field and

$$\hat{K}_{n,m} = \begin{Vmatrix} K_{s1}\pm K_{i1} & \frac{K_{s2}\pm K_{i2}}{2} & 0 \\ \frac{K_{s2}\pm K_{i2}}{2} & 0 & 0 \\ 0 & 0 & 0 \end{Vmatrix} = K_{n,m}\begin{Vmatrix} 1 & \frac{1}{2}a_0\kappa & 0 \\ \frac{1}{2}a_0\kappa & 0 & 0 \\ 0 & 0 & 0 \end{Vmatrix}, \quad K_{n,m} = K_{s1}\pm K_{i1}, \quad \hat{\lambda} = \begin{Vmatrix} \lambda+\lambda_{\text{T}} & 0 & 0 \\ 0 & \lambda & 0 \\ 0 & 0 & \lambda \end{Vmatrix}. \tag{S21}$$

In the following, we derive the $\sigma$-model. We assume that $m \ll n \sim 1$ and perform our analysis in the long-wave approximation (slow temporal and spatial variation of $m$ and $n$). These assumptions are valid for weak gradients of the order parameters and when external magnetic fields are sufficiently small. The effective action of the system $\mathscr{A} = \int \mathscr{L} ds dt$ with constraints (S11) reads

$$\mathscr{A}^{\text{eff}} = \mathscr{A} + \varsigma_1 \int (m \cdot n) ds dt + \varsigma_2 \int (m^2 + n^2 - 1) ds dt, \tag{S22}$$

where $\varsigma_{1,2}$ are the Lagrange multipliers. The condition of extremum of $\mathscr{A}$ reads

$$\frac{2M_\text{S}}{\gamma_0}[n \times \dot{n}] + \frac{\delta E}{\delta m} = n\left(n \cdot \frac{\delta E}{\delta m}\right) + n\left(m \cdot \frac{\delta E}{\delta n}\right) + m\left(n \cdot \frac{\delta E}{\delta n}\right) \tag{S23}$$

where $E = \int \mathscr{W} ds$ is the effective energy and the terms quadratic with respect to $m$ and its derivatives are omitted. Taking into account smallness of the spatial and temporal derivatives of $m$, the vector of ferromagnetism reads

$$m = -n \times \left[\frac{M_\text{S}}{\Lambda \gamma_0}\dot{n} + \frac{\lambda n' - 2M_\text{S} H}{2\Lambda} \times n\right] + \mathcal{O}(\varsigma). \tag{S24}$$

Here, $\varsigma = a_0/(2\ell)$ is the expansion parameter, which determines the scale of the magnetization. Then, the Lagrangian density $\mathscr{L}$ reads

$$\mathscr{L} = \frac{M_\text{S}^2}{\Lambda \gamma_0^2}|\dot{n}|^2 - \underbrace{An'^2}_{A\sum_\nu n_\nu'^2 + E_\text{x}^\text{A} + E_\text{x}^\text{DM}} + n\hat{K}_n n - 2\frac{M_\text{S}^2}{\Lambda \gamma_0}H \cdot [n \times \dot{n}] + \frac{M_\text{S}^2}{\Lambda}[H \times n]^2. \tag{S25}$$

Here, $A = A_0 - \lambda^2/(4\Lambda) = A_0/2$ and the exchange term in the curvilinear reference frame manifests the anisotropic and chiral contributions as well [5]. For classical systems, we can omit the so-called topological term [1, 6] with the following contribution to the action:

$$\mathscr{A}^{\text{topo}} = -\frac{\lambda M_\text{S}}{\Lambda \gamma_0} \int n' \cdot [n \times \dot{n}] ds dt = 2\pi \hbar S Q, \tag{S26}$$

where $Q = \frac{1}{4\pi}\int n \cdot [n' \times \dot{n}] ds dt \in \mathbb{Z}$ is the topological invariant (skyrmion number) of the mapping of the $(s,t)$ plane onto the sphere $n^2 = 1$.

Units for a straight chain with the intrinsic easy-axis anisotropy can be rescaled in terms of the exchange and anisotropy fields $H_\text{x} = 4\mathscr{J}S/(g\mu_\text{B})$ and $H_\text{a} = K_n/M_\text{S} = (\mathscr{K}_\text{SI} + \mathscr{K}_\text{II})S/(g\mu_\text{B})$. Then, the spin-flop field, defined for the case when the system possesses a micromagnetic easy axis of anisotropy, reads $H_\text{sf} = \sqrt{H_\text{x} H_\text{a}} = 2S\sqrt{\mathscr{J}(\mathscr{K}_\text{SI} + \mathscr{K}_\text{II})}/(g\mu_\text{B})$ and the frequency of the antiferromagnetic resonanse $\omega_0 = \gamma_0\sqrt{H_\text{x} H_\text{a}} = 2S\sqrt{\mathscr{J}(\mathscr{K}_\text{SI} + \mathscr{K}_\text{II})}/\hbar$. Note, that the definition of the magnetic length $\ell = \sqrt{A/K_n}$ coincides with (S5) and $\Lambda = H_\text{x} M_\text{S}$. Then, we can rescale units in (S24) and (S25) using Table S2:

$$\begin{aligned}\mathscr{\bar L} &= (\partial_\tau n - [h \times n])^2 - (\partial_\xi n)^2 + n_\text{T}^2 + 2\varkappa \varsigma n_\text{T} n_\text{N},\\ m &\approx -\varsigma n \times [\partial_\tau n + (\partial_\xi n - h) \times n].\end{aligned} \tag{S27}$$

where $\mathscr{\bar L} = \mathscr{L}/(H_\text{a} M_\text{S})$ is the normalized Lagrangian density, $\tau = \omega_0 t$ is the normalized time, $\xi = s/\ell$ is the normalized coordinate along $\gamma$ and $h = H/H_\text{sf}$.

## IV. SPIN-LATTICE SIMULATIONS

Numerical simulations of AFM spin chains are performed for flat rings using spin-lattice simulator SLaSi [7]. The ring is represented by a spin chain of discrete unit magnetic moments $\mu_i$, $i = \overline{1,N}$, with periodic boundary conditions. The number of magnetic sites $N$ determines the ring curvature for the given $a_0$. We determine the equilibrium magnetic states by solving the system of coupled Landau–Lifshitz–Gilbert equations [8, 9] for magnetic moments

$$\frac{d\mu_i}{dt} = \gamma_0 \mu_i \times \frac{\partial \mathscr{H}}{\partial \mu_i} + \alpha_\text{G} \mu_i \times \frac{d\mu_i}{dt} \tag{S28}$$

for the Hamiltonian $\mathscr{H} = \mathscr{H}_\text{x} + \mathscr{H}_\text{SI} - 2\mu_\text{B} S \sum_i \mu_i \cdot H$, where the last term represents the interaction with the external magnetic field and $\alpha_\text{G}$ is the Gilbert damping. The overdamped dynamics with $\alpha_\text{G} = 0.5$ is simulated since we are interested in the equilibrium states only. The equilibrium magnetic state is determined after relaxation up to $|d\mu_i/d\tau| < 10^{-38}$.

| Quantity | Notation | Unit of measurement | Value in spin lattice units |
|---|---|---|---|
| Exchange field | $H_x$ | | $\dfrac{4\mathscr{J}S}{g\mu_B}$ |
| Anisotropy field | $H_a$ | $\dfrac{K_n}{M_S}$ | $\dfrac{(\mathscr{K}_{SI}+\mathscr{K}_{II})S}{g\mu_B}$ |
| Time | $\tau = \omega_0 t$ | $\omega_0 = \gamma_0\sqrt{H_x H_a}$ | $\dfrac{2S}{\hbar}\sqrt{\mathscr{J}(\mathscr{K}_{SI}+\mathscr{K}_{II})}$ |
| Length | $\xi = \dfrac{r}{\ell}$ | $\ell = \sqrt{\dfrac{A}{K_n}}$ | $a_0\sqrt{\dfrac{\mathscr{J}}{|\mathscr{K}_{SI}+\mathscr{K}_{II}|}}$ |
| Field | $h = \dfrac{H}{H_{sf}}$ | $H_{sf} = \sqrt{H_x H_a}$ | $\dfrac{2S}{g\mu_B}\sqrt{\mathscr{J}(\mathscr{K}_{SI}+\mathscr{K}_{II})}$ |
| Magnetization | $M_S$ | | $\dfrac{g\mu_B S}{2a_0}$ |
| Expansion parameter | $\varsigma$ | $\sqrt{\dfrac{H_a}{H_x}}$ | $\dfrac{a_0}{2\ell}$ |

TABLE S2: Scaling of units.

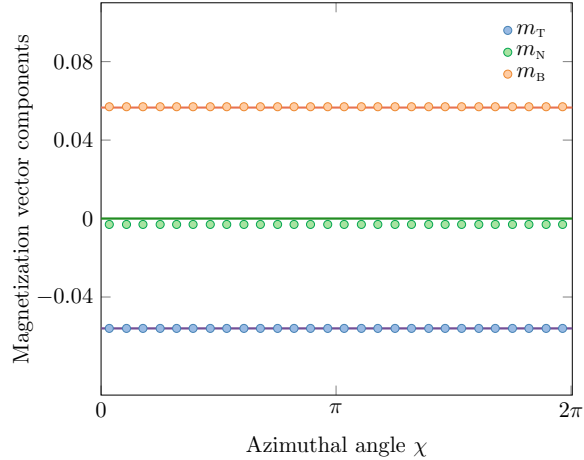

FIG. S1: **Magnetization of an AFM ring above the spin-flop transition.** Components of the magnetization in the TNB reference frame. Solid lines correspond to analytics (Eq. (9) in the main text). Symbols are the results of the spin-lattice simulation ($\ell = 5a_0$, $\varkappa = 0.56$, $\varsigma = 0.1$, $h = 0.56$).

We find that in a strong magnetic field applied along the ring axis, the equilibrium direction of the magnetic moments is mainly along $e_N$. The tilt angle of the Néel and magnetization vectors $\psi$ remains constant for the given ring curvature in a wide range of external fields between $H_{sf}$ and $H_x$ fields (Fig. 3b in the main text, other parameters except the ring curvature are the same as in Fig. S1). The vector of ferromagnetism in equilibrium is shown in Fig. S1, where the coordinate along $\gamma$ is measured in $\chi = s/R$.

## V. MAGNETIZATION OF AFM TEXTURES IN CURVILINEAR SPIN CHAINS

In contrast to the case of two-dimensional [10, 11] and three-dimensional [12] AFMs, non-collinear AFM textures in spin chains possess an intrinsic magnetization related to the translational invariance breaking with respect to the exchange of the AFM sublattices [13–15]. This results in the appearance of the parity breaking term with the coefficient $\lambda$ in the exchange energy (1). The illustration of this effect is shown in Fig. S2, see also the discussion and comparison with the Haldane's mapping in Refs. [13–15]. In 1D, the ground states of a classical AFM spin chain with the opposite directions of the sublattice

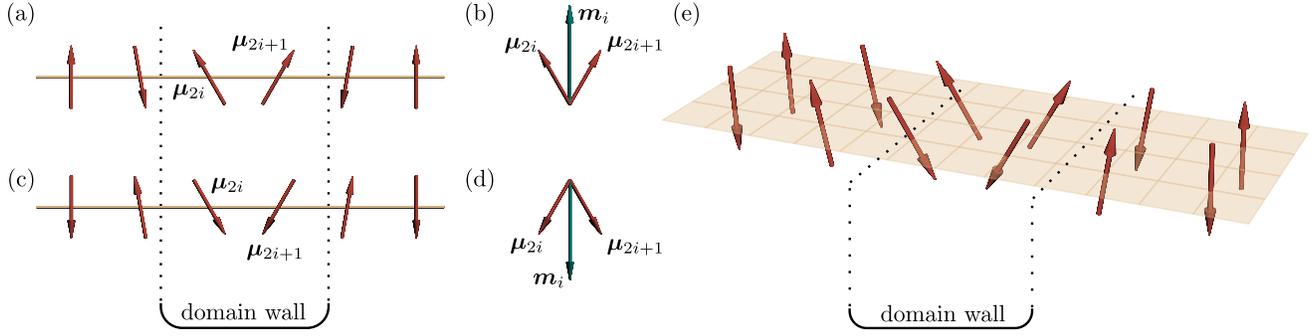

FIG. S2: **AFM domain walls in spin chains and 2D lattices.** (a) Schematics of a domain wall in a spin chain and (b) the corresponding magnetic moment at the domain wall. (c,d) The same for the opposite direction of the sublattices. (e) Schematics of the domain wall in a 2D G-type AFM. The lattice planes perpendicular to the domain wall possess opposite intrinsic magnetic moments and compensate each other.

magnetizations can be transformed in each other by a translation by one lattice constant. The presence of a non-collinear texture breaks this symmetry, see Fig. S2(a,c). The nearest-neighbor exchange connects the rotation of both sublattices in a smooth manner, which leads to a non-zero magnetic moment of the texture [Fig. S2(b,d)]. The direction of this moment is determined by the direction of the magnetic moments in each of the sublattices in the ground state. This makes both domain wall states to be energetically equivalent. The magnetic moment of the AFM texture does not appear in two- and three-dimensional cases for infinite G-type AFMs because of the compensation by the neighboring lattice planes, see Fig. S2(e).

For a curvilinear AFM spin chain, the strictly homogeneous texture in the laboratory reference frame is forbidden due to the competition between the exchange interaction and anisotropy, which follows the chain geometry [5]. Thus, even in the first order in $\varsigma$, there is a non-zero magnetization even for a locally homogeneous AFM texture in (S24), produced by the term $\boldsymbol{n}'$. This is the source of the tangential magnetization in an AFM ring, shown in Fig. 3c of the main text and in Fig. S1. The direction $m_\mathrm{T}$ can be changed to the opposite one by the change of the in-plane direction of $\boldsymbol{\mu}_{\mathrm{TN}i} \to -\boldsymbol{\mu}_{\mathrm{TN}i}$, while the energy of the system remains the same.

[1] B. A. Ivanov and A. K. Kolezhuk, Solitons in low-dimensional antiferromagnets, Low Temperature Physics **21**, 275 (1995).
[2] E. Manousakis, The spin-1/2 Heisenberg antiferromagnet on a square lattice and its application to the cuprous oxides, Reviews of Modern Physics **63**, 1 (1991).
[3] P. W. Anderson, New approach to the theory of superexchange interactions, Physical Review **115**, 2 (1959).
[4] N. Nagaosa, *Quantum Field Theory in Strongly Correlated Electronic Systems (Theoretical and Mathematical Physics)* (Springer, 1999).
[5] O. V. Pylypovskyi, D. Y. Kononenko, K. V. Yershov, U. K. Rößler, A. V. Tomilo, J. Fassbender, J. van den Brink, D. Makarov, and D. D. Sheka, Curvilinear one-dimensional antiferromagnets, Nano Letters **20**, 8157 (2020).
[6] B. A. Ivanov, Mesoscopic antiferromagnets: statics, dynamics, and quantum tunneling (Review), Low Temperature Physics **31**, 635 (2005).
[7] SLaSi spin–lattice simulations package.
[8] L. D. Landau and E. M. Lifshitz, On the theory of the dispersion of magnetic permeability in ferromagnetic bodies, Phys. Zs. Sowjet.; Reproduces in in Ukr. J. Phys. 53 25-35 (2008) **8**, 153 (1935).
[9] T. Gilbert, A phenomenological theory of damping in ferromagnetic materials, IEEE Transactions on Magnetics **40**, 3443 (2004).
[10] S. Komineas and N. Papanicolaou, Vortex dynamics in two-dimensional antiferromagnets, Nonlinearity **11**, 265 (1998).
[11] S. Komineas and N. Papanicolaou, Traveling skyrmions in chiral antiferromagnets, SciPost Physics **8**, 10.21468/scipostphys.8.6.086 (2020).
[12] O. V. Pylypovskyi, A. V. Tomilo, D. D. Sheka, J. Fassbender, and D. Makarov, Boundary conditions for the néel order parameter in a chiral antiferromagnetic slab, ArXiv e-prints (2021), 2101.09134.
[13] N. Papanicolaou, Antiferromagnetic domain walls, Physical Review B **51**, 15062 (1995).
[14] N. Papanicolaou, Dynamics of domain walls in weak ferromagnets, Physical Review B **55**, 12290 (1997).
[15] E. G. Tveten, T. Müller, J. Linder, and A. Brataas, Intrinsic magnetization of antiferromagnetic textures, Physical Review B **93**, 104408 (2016).


	
\end{document}